\documentclass[twocolumn]{aa}

\usepackage{graphicx}
\usepackage{txfonts}
\usepackage{natbib}
\bibpunct{(}{)}{;}{a}{}{,}   

\begin{document}

\titlerunning{asteroseismic test of diffusion theory}
\authorrunning{Metcalfe et al.}

\title{An asteroseismic test of diffusion theory in white dwarfs} 

\author{T.~S.~Metcalfe\inst{1}
\and R.~E.~Nather\inst{2}
\and T.~K.~Watson\inst{3}
\and S.-L.~Kim\inst{4}
\and B.-G.~Park\inst{4}
\and G.~Handler\inst{5}}

\institute{High Altitude Observatory, National Center for Atmospheric 
     Research, P.O. Box 3000, Boulder, CO 80307-3000 USA
\and Department of Astronomy, University of Texas, Austin, TX 78712 USA
\and Southwestern University, 1001 E. University Avenue, Georgetown, 
     TX 78626 USA
\and Korea Astronomy Observatory, Daejeon, 305-348, Korea
\and Institut f{\"u}r Astronomie, Universit{\"a}t Wien, 
     T{\"u}rkenschanzstra{\ss}e 17, 1180 Wien, Austria}

\date{Received 2004 Dec 23; accepted 2005 Feb 06}

\abstract{The helium-atmosphere (DB) white dwarfs are commonly thought to
be the descendants of the hotter PG~1159 stars, which initially have
uniform He/C/O atmospheres. In this evolutionary scenario, diffusion
builds a pure He surface layer which gradually thickens as the star cools.
In the temperature range of the pulsating DB white dwarfs ($T_{\rm eff}
\sim 25\,000$~K) this transformation is still taking place, allowing
asteroseismic tests of the theory. We have obtained dual-site observations
of the pulsating DB star CBS~114, to complement existing observations of
the slightly cooler star GD~358. We recover the 7 independent pulsation
modes that were previously known, and we discover 4 new ones to provide
additional constraints on the models. We perform objective global fitting
of our updated double-layered envelope models to both sets of
observations, leading to determinations of the envelope masses and pure He
surface layers that qualitatively agree with the expectations of diffusion
theory. These results provide new asteroseismic evidence supporting one of
the central assumptions of spectral evolution theory, linking the DB white
dwarfs to PG~1159 stars.

}

\maketitle

\section {Introduction}

The origin and evolution of the helium-atmosphere (DB) white dwarf stars
has long been a subject of debate. They comprise roughly 20\% of the
population of field white dwarfs, with most of the remaining 80\%
consisting of their hydrogen-atmosphere (DA) cousins. The majority of
white dwarfs in both of these spectral classes are thought to arise from
the evolution of isolated main-sequence stars with masses that are
insufficient to ignite carbon fusion, which ultimately leave their hot
carbon/oxygen cores to descend the white dwarf cooling track. The
bifurcation into two spectral classes is thought to occur during
post-asymptotic-giant-branch (post-AGB) evolution when, in some cases, a
very late thermal pulse burns off the residual hydrogen in the envelope,
producing a nearly pure helium atmosphere \citep{iben83}. Such objects are
then supposed to return to the white dwarf cooling track as PG~1159 stars,
which are widely believed to be the precursors of most DB white dwarfs.

The primary difficulty with this scenario is the paucity of DB white
dwarfs with temperatures between 45\,000 and 30\,000~K---a phenomenon
known as the ``DB gap'' \citep{lie86}, which persists even in the latest
results from the Sloan Digital Sky Survey \citep{kle04}. If the PG~1159
stars and DB white dwarfs are linked, then why don't we see the long-lived
intermediate phase? The {\it spectral evolution theory} advanced by
\cite{fw87,fw97} postulates that traces of hydrogen ($M_H \sim
10^{-15}~M_*$) in the envelopes of hot PG~1159 stars float to the surface
and disguise them as DA stars within this temperature interval. A growing
helium convection zone eventually dilutes the hydrogen in the photosphere
until it is no longer detectable at 30\,000~K, and the star continues its
evolution as a DB. A relative overabundance of DA stars inside the DB gap
supports this hypothesis \citep[see][ their Fig.~11]{kle04}. But lower
than predicted surface hydrogen abundances, measured for several DB stars
just below the gap, remain a problem for spectral evolution theory
\citep{pro00}.

If we assume that there {\it is} an evolutionary connection between the
PG~1159 stars and the cooler DB white dwarfs, we can look to several
independent groups who have used time-dependent diffusion calculations to
follow the changes in the interior structure of these objects as they cool
\citep{dk95,fb02,ac04}. The hot PG~1159 stars, having recently emerged
from the born-again phase, contain envelopes with a nearly uniform mixture
of helium (He), carbon (C), and oxygen (O) out to the photosphere
\citep{dh98,her99}. As they cool, the He diffuses upward and gradually
accumulates to form a chemically pure surface layer. This leads to a
double-layered structure, with the pure He surface layer overlying the
remainder of the uniform He/C/O envelope, all above the degenerate C/O
core.

A key prediction of the diffusion models is that, for a given stellar
mass, the pure He surface layer will steadily grow thicker as the DB star
cools. The only available observational tests of this prediction come from
asteroseismology---the study of the internal structure of stars through
the interpretation of their pulsation periods. Fortunately, non-radial
oscillations are naturally excited in otherwise normal DB white dwarf
stars as they cool through a narrow range of temperatures near 25\,000~K.
However, there are only 9 such pulsating DB stars (DBVs) presently known,
with effective temperatures between 22\,400 and 28\,400~K \citep{bea99}.

The most thoroughly studied DBV star is GD~358, which has been the target
of three coordinated observing campaigns of the Whole Earth Telescope
\citep[WET;][]{nat90,win94,vui00,kep03}. These observations revealed
dozens of independent pulsation periods, including a sequence of 11 dipole
\citep[$\ell$=1,~$m$=0;][]{kot03} modes of consecutive radial overtone
\citep[$k$;][]{bw94}, each providing a complementary probe of the stellar
interior. These data formed the basis for the only asteroseismic tests of
the diffusion models to date \citep{dk95,fb02}. In the most extensive of
these tests, \cite{fb02} produced a targeted grid of double-layered
evolutionary models to try to fit the periods of GD~358 in detail. They
found evidence of two composition transition zones near outer mass
fractions of $10^{-5.8}$ and $10^{-3.0}~M_*$, which they interpreted as
the base of the pure He surface layer and the still-uniform He/C/O
envelope, respectively.  However, \cite{mmw03} showed that there is a
potential ambiguity in the inferred locations of these internal
structures, due to an inherent core-envelope symmetry for models of DBV
stars. As a consequence, real structure at specific locations in the core
could be mis-interpreted as structure at symmetric points in the
envelope, and vice versa.

The best way of circumventing this potential ambiguity is to fit
additional stars using the same models, to determine whether the
independent observations are consistent with the same underlying physical
processes. Such model-fitting attempts for DBV stars other than GD~358
have been hampered by the lack of suitable observational data.  
\citeauthor{hmw02} (\citeyear{hmw02}, hereafter HMW) found a total of 7
independent pulsation periods in the slightly hotter DBV star CBS~114, and
\cite{mmk03} attempted to fit these observations using double-layered
envelope models that were also applied to GD~358. They found qualitative
agreement with the predictions of diffusion theory (thinner surface He
layers for hotter white dwarfs of comparable mass) when using models that
also included an adjustable C/O profile in the core. However, the large
number of free parameters in their model, relative to the number of
observed periods in CBS~114, made these results somewhat uncertain.

\section{Observations \& Data Reduction}

Motivated by the need for additional pulsation periods to constrain the
model-fitting of CBS~114, we organized a dual-site campaign using 2\,m
class telescopes in February 2004. Our primary goals were to confirm the
periods of the 7 pulsation modes found by HMW, and to discover additional
modes below the $\sim$5 milli-magnitude noise limit of the previous
observations (obtained over 3 weeks from the 0.75\,m telescope at SAAO).  
This campaign was awarded 7 nights (2004 Feb 19-25) on the 1.8\,m
telescope at Bohyunsan Optical Astronomy Observatory (BOAO) in Korea, and
7 nights (2004 Feb 21-27) on the 2.1\,m telescope at McDonald Observatory
(McD) in Texas.

At BOAO we used a thinned SITe 2K CCD camera with no filter. To minimize
the readout time, we binned the pixels 2$\times$2 for an effective
resolution of 0.68 arcseconds~pixel$^{-1}$, and we used only the central
portion of the detector for a field of view $\sim$4 arcminutes on a side.
On the first night we used an exposure time of 8 seconds for a cycle time
of 20 seconds, but on subsequent nights we increased the exposure time to
20.5 seconds for a cycle time of 30 seconds. This system has a gain of
1.8~$e^- {\rm ADU}^{-1}$ and a readout noise of 7~$e^-$ rms.

The observations at McD were obtained using the Argos time-series CCD
photometer \citep{nm04} with no filter. This instrument uses a
back-illuminated 512$\times$512 frame-transfer CCD mounted at the f/3.9
prime focus, for a field of view 2.8 arcminutes on a side and a resolution
of 0.33 arcseconds~pixel$^{-1}$. We used an exposure time of 10 seconds,
which was also the cycle time---the transfer of each image to the masked
on-chip buffer requires only 310~$\mu$s, and the 280~ms readout takes
place during the next exposure. This system has a gain of 2~$e^- {\rm
ADU}^{-1}$ and a readout noise of 8~$e^-$ rms. The observations collected
for this dual-site campaign are summarized in Table~\ref{tab1}.

\begin{table}
\begin{center}
\caption{Journal of observations.\label{tab1}}
\begin{tabular}{llccc}
\hline
Run     & Telescope & Date        &  Start   & Length \\
        &           &             &   (UT)   & (s)    \\
\hline
 040219 & BOAO 1.8m & 2004 Feb 19 & 11:15:50 & 33540  \\
 040220 & BOAO 1.8m & 2004 Feb 20 & 10:58:19 & 13080  \\
  A0844 & McD 2.1m  & 2004 Feb 21 & 04:17:05 & 11050  \\
 040223 & BOAO 1.8m & 2004 Feb 23 & 10:42:57 & 29520  \\
 040224 & BOAO 1.8m & 2004 Feb 24 & 15:58:30 & 14520  \\
 040225 & BOAO 1.8m & 2004 Feb 25 & 12:49:33 & 24840  \\
  A0847 & McD 2.1m  & 2004 Feb 26 & 01:53:07 & 36120  \\
  A0848 & McD 2.1m  & 2004 Feb 27 & 01:51:17 & 38550  \\
\hline
\end{tabular}
\end{center}
\end{table}


\begin{figure*}
\centering\includegraphics[angle=270,width=18cm]{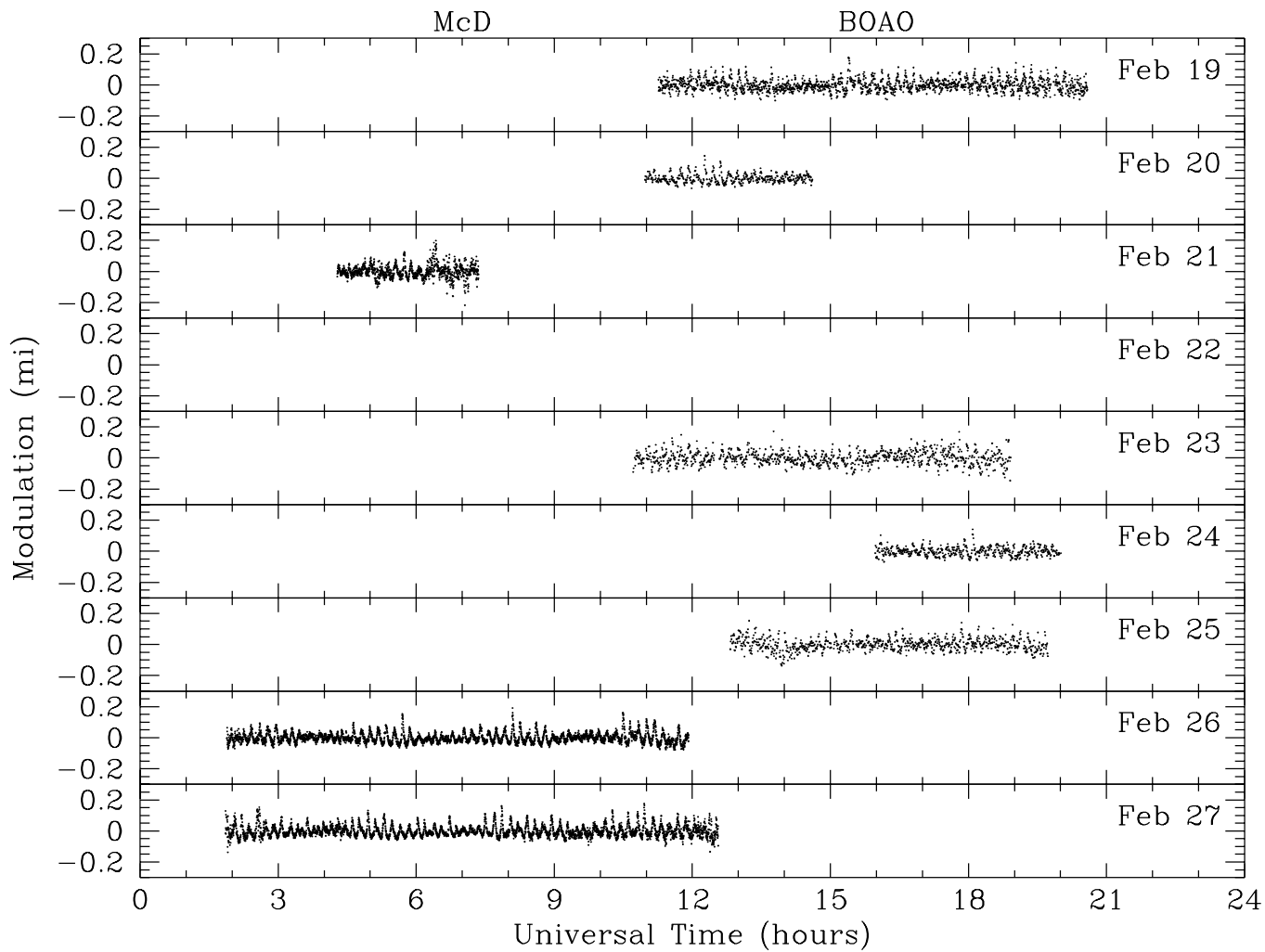}
\caption{Reduced light curves for CBS~114 during our dual-site 
campaign in 2004, showing fractional intensity variations of
$\pm$15 percent with a dominant period of $\sim$650 seconds.}
\label{fig1}
\vspace*{10pt}
\end{figure*}

The raw CCD frames were analyzed on-the-fly during the observing campaign
to monitor the data quality, but final reductions were done uniformly
using the IRAF high-speed photometry (HSP) package developed by Antonio
Kanaan (personal communication). After correcting each image for bias
level and flat field effects, we extracted aperture photometry for CBS~114
and three nearby comparison stars. The aperture was typically set to have
a radius between 2 and 4 arcseconds with a surrounding annulus to
determine the local sky level, but the apertures were adjusted for each
data set to maximize the S/N of the resulting light curves. The
time-series measurements were then imported into XQED \citep{rid03}, the
photometry analysis software for the WET. For each light curve, we
subtracted the local sky contribution and performed a point-by-point
division of the target star by the average of the three comparison stars.  
We then corrected for any low-frequency variations (e.g., caused by
differential color extinction) by fitting and removing a second-order
polynomial. Finally, we subtracted the mean light level and applied
barycentric timing corrections to the exposure mid-times. The resulting
light curves for CBS~114 are shown in Fig.~\ref{fig1}.

\section{Frequency Analysis}

\begin{figure*}
\centering\includegraphics[angle=270,width=18cm]{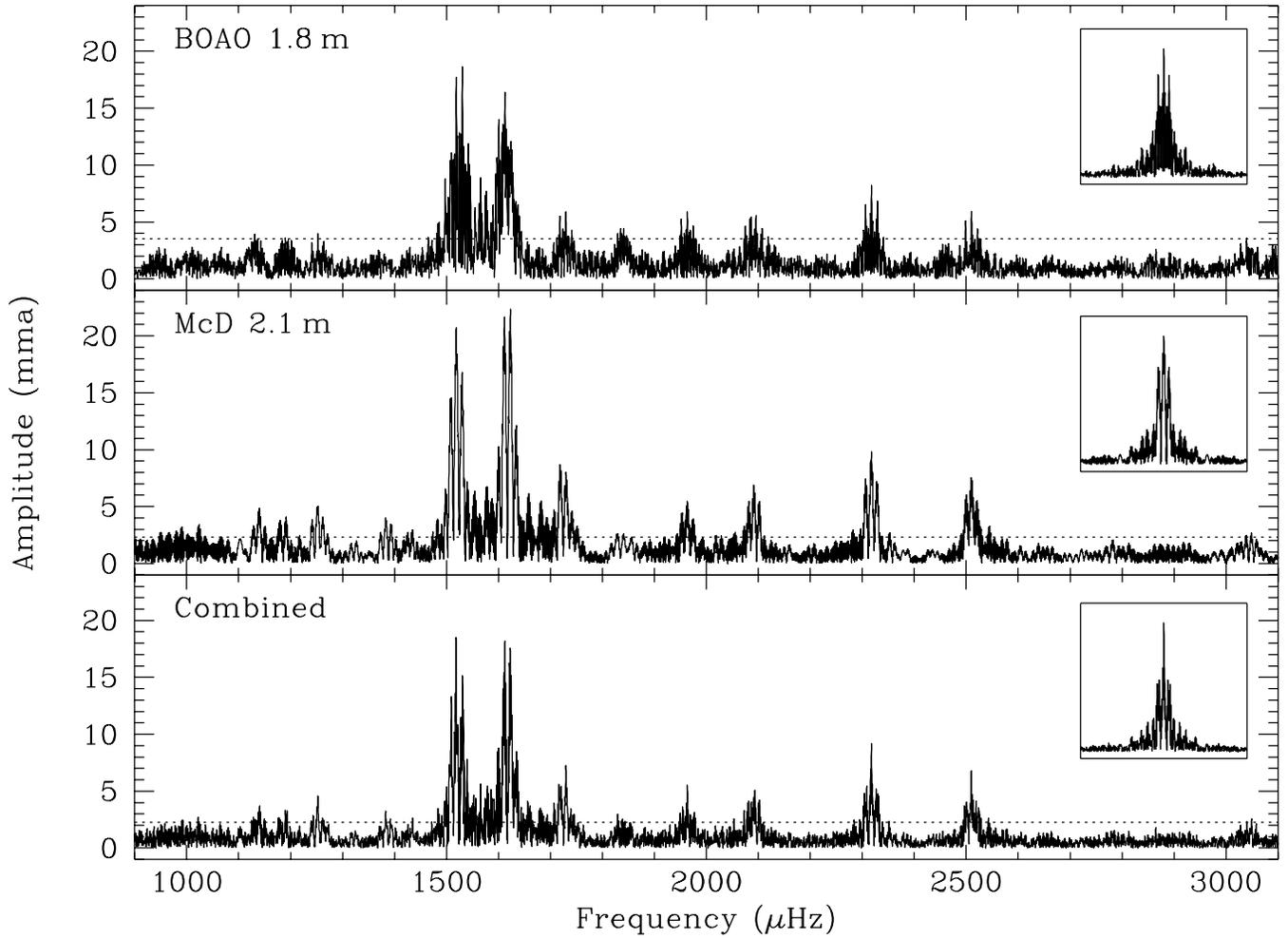}
\caption{Fourier Transforms (FTs) of the light curves shown in
Fig.~\ref{fig1} over the range of independent frequencies exhibited by
CBS~114. The three panels show the FTs for the data from BOAO (top), McD
(middle), and for the combined data set (bottom), along with the
corresponding window functions (inset) and significance cuts (dotted
line).}
\label{fig2} 
\end{figure*}

The Fourier Transforms (FTs) of these light curves, for each observatory
and for the combined data set, are shown in Fig.~\ref{fig2}. The
corresponding window functions---the FT of a single sinusoid sampled in
the same way as the data---are shown to the same scale in each panel. The
distributed sampling of the BOAO data yields better effective frequency
resolution, while the slightly larger aperture and higher cadence of the
McD data leads to a higher S/N. The full data set combines all of these
strengths, and reduces the amplitude of the $1\,d$ aliases (separated from
the main frequency by $\pm11.6~\mu$Hz).

To determine the significance level of peaks in each of these FTs, we
followed the analysis of \cite{kep93}. Since power is the square of the
amplitude, the average power out to the Nyquist frequency is a measure of
the mean-square deviation from zero amplitude, which is dominated by the
noise since the signal is confined to a relatively narrow range of
frequencies. The square-root of this value is the root-mean-square
amplitude of the noise, which is analogous to the standard deviation
($\sigma_{\rm noise}$). By adopting a significance cut at $3\sigma_{\rm
noise}$, we ensure that only $\sim$0.3\% of the peaks with this amplitude
will not be real signals (assuming the noise is Gaussian). Peaks with
amplitudes much above the significance cut, and those detected in multiple
independent data sets, have an even lower chance of being due to noise.
The $3\sigma_{\rm noise}$ significance level is indicated by the dotted
line in each panel of Fig.~\ref{fig2}, which is at 3.52, 2.33, and
2.26~mma for the BOAO, McD, and combined data sets respectively. For the 
range of frequencies shown, we expect only $\sim$4 noise peaks to appear 
above these cuts.

We performed a frequency analysis of the combined data set using the
program PERIOD98 \citep{spe98}. We started by iteratively fitting and
subtracting the frequencies with the largest amplitudes, refining our
estimates of the period, amplitude, and phase for each signal using
simultaneous multiple sine-wave fitting. This allowed us to recover the 7
independent pulsation frequencies found by HMW, with a few important
differences. We resolved both of the two largest amplitude modes near 1520
and 1610~$\mu$Hz into triplets, though the splitting is close to the
$1\,d$ aliases (perhaps explaining why the earlier single-site data failed
to detect them, despite a longer time baseline). We also detected an
additional component in the 1520~$\mu$Hz multiplet at 1514.1~$\mu$Hz,
which could be caused by amplitude and/or frequency modulation during the
observations or might be caused by higher-order combination frequencies,
comparable to the ``$k=\hat{15}$'' mode seen in GD~358 in 1994
\citep{vui00}. Our frequencies for the modes near 1830 and 1960~$\mu$Hz
are both systematically lower than in HMW, but by less than would be
expected from $1\,d$ alias problems. This might be the result of an
underlying multiplet structure with intrinsic amplitude modulation among
the various components \citep[e.g., see][ their Fig.~2]{rnm76}. We also
found that our mode near 2300~$\mu$Hz was the $+1\,d$ alias of the
frequency found by HMW---a possibility that they could not rule out.
Finally, we resolved the mode near 2510~$\mu$Hz, as well as a newly
discovered mode near 2090~$\mu$Hz, into doublets with a splitting
comparable to the frequency shifts seen in the 1830 and 1960~$\mu$Hz
modes.

In addition to recovering the 7 pulsation modes found by HMW, we also
discovered 4 new ones. The largest is the doublet near 2090~$\mu$Hz, which
nicely fills the gap in the sequence of consecutive radial overtones
identified by HMW. A marginal peak near this frequency also appeared in
the discovery data of \cite{wc88,wc89}, but it was below the detection
threshold of HMW. The three other new modes appear at frequencies near
1190, 1250, and 1380~$\mu$Hz. The period spacing of these peaks suggests
that they are additional independent pulsation modes with higher radial
overtones than those previously seen. Two additional frequencies had
amplitudes near our significance threshold. These could tentatively be
identified as an even higher radial overtone near 1140~$\mu$Hz, and a
possible rotationally split component of the mode near 1720~$\mu$Hz. These
marginal detections prompted us to compute weighted FTs, following the
suggestion of \cite{han03}. The results were nearly identical to the
unweighted case, and the marginal nature of these two frequencies was
unchanged.

\begin{figure*}
\centering\includegraphics[angle=270,width=18cm]{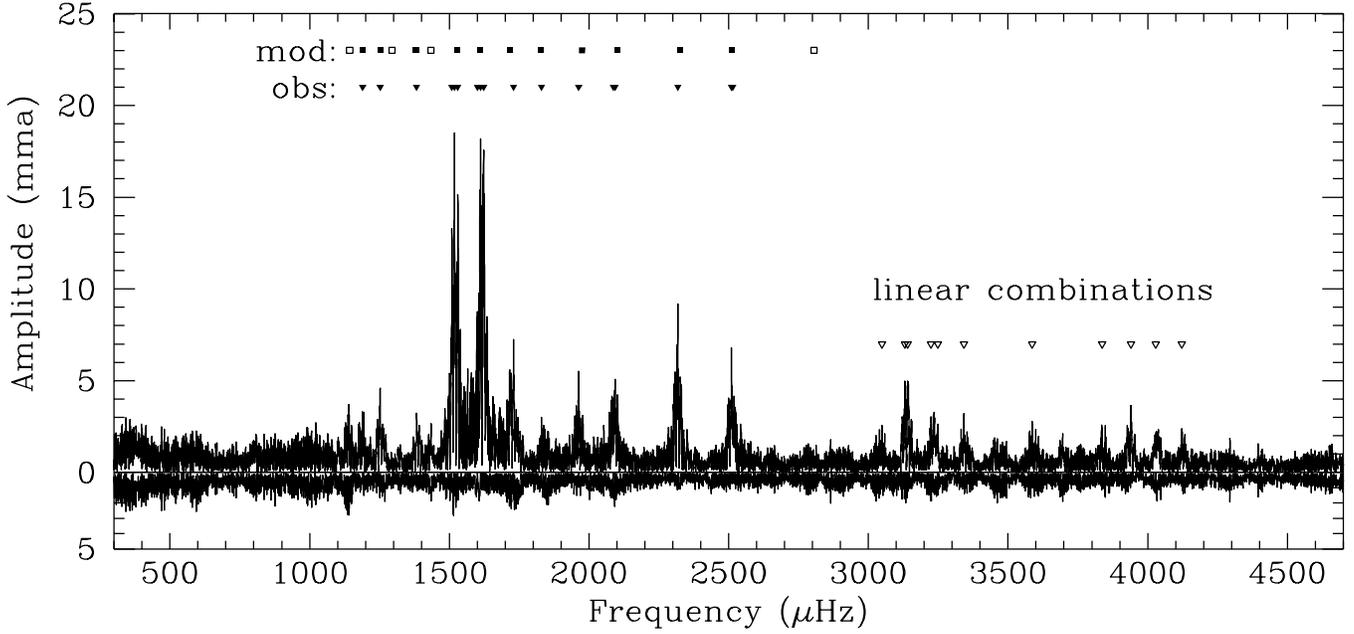}
\caption{The Fourier Transform of our combined data set for CBS~114 before
and after removing the 28 sinusoids listed in Table~\ref{tab2}. The 
independent and linear combination frequencies are indicated by the solid 
and open triangles respectively. The frequencies of our optimal model 
matched to the observed modes are shown as solid squares, while open 
squares denote model frequencies with no detected counterpart.}
\label{fig3}
\end{figure*}

Aside from the independent pulsation modes observed in CBS~114, we also
detected 11 significant linear combination frequencies. Fundamentally,
these frequencies arise in our analysis because the Fourier Transform
decomposes the signal into sinusoidal components. A casual inspection of
the light curves shown in Fig.~\ref{fig1} reveals that they are not
perfectly sinusoidal. Instead, they tend to have sharp peaks and shallow
troughs, which is commonly interpreted to be associated with a nonlinear
response of the surface convection zone to the pulsations \citep[e.g.,
see][]{bri92,wu01,ik01,mon05}. The earlier observations by HMW revealed
two such frequencies, both of which are among those we detected. The
combination frequencies identified in our data involve the six modes or
multiplets with the largest amplitudes.

Our final simultaneous fit of 28 sinusoids to the data using PERIOD98
contains a total of 17 independent modes and 11 linear combination
frequencies, which are all listed in Table~\ref{tab2}. In Fig.~\ref{fig3}
we show the FT of our combined data set before and after the removal of
these 28 frequencies (indicated by solid and open triangles for the
independent and combination frequencies, respectively). The marginal
signals near 1140, 1514, and 1720~$\mu$Hz are apparent in the residuals.

\begin{table}
\begin{center}
\caption{The multifrequency solution for CBS~114. Also shown is
the mode identification from our optimal model.\label{tab2}}
\begin{tabular}{lcccrrr}
\hline
Sinusoid   & Frequency & Period &$\left<{\rm Amp}\right>$ & 
                              \multicolumn{3}{c}{Identification}\\
           & ($\mu$Hz) &  (s)   &  (mmag) & $k$ & $\ell$ & $m$  \\
\hline
 $f_{1}\dotfill$ & 1189.9 & 840.38 &  3.1 & 20  &   1    & 0    \\ 
 $f_{2}\dotfill$ & 1252.4 & 798.46 &  4.3 & 19  &   1    & 0    \\ 
 $f_{3}\dotfill$ & 1383.0 & 723.06 &  3.0 & 17  &   1    & 0    \\ 
 $f_{4}\dotfill$ & 1509.0 & 662.67 &  7.5 & 15  &   1    & $-$1 \\ 
 $f_{5}\dotfill$ & 1518.8 & 658.44 & 11.8 & 15  &   1    & 0    \\ 
 $f_{6}\dotfill$ & 1530.7 & 653.30 &  9.0 & 15  &   1    & $+$1 \\ 
 $f_{7}\dotfill$ & 1601.2 & 624.55 &  4.4 & 14  &   1    & $-$1 \\
 $f_{8}\dotfill$ & 1612.1 & 620.31 & 15.4 & 14  &   1    & 0    \\ 
 $f_{9}\dotfill$ & 1622.7 & 616.27 & 13.3 & 14  &   1    & $+$1 \\ 
$f_{10}\dotfill$ & 1729.8 & 578.09 &  7.0 & 13  &   1    & 0    \\ 
$f_{11}\dotfill$ & 1829.6 & 546.56 &  3.0 & 12  &   1    & 0    \\ 
$f_{12}\dotfill$ & 1963.2 & 509.37 &  5.5 & 11  &   1    & 0    \\ 
$f_{13}\dotfill$ & 2086.5 & 479.27 &  4.8 & 10  &   1    & ?    \\ 
$f_{14}\dotfill$ & 2093.6 & 477.65 &  4.9 & 10  &   1    & ?    \\ 
$f_{15}\dotfill$ & 2317.7 & 431.45 &  8.9 &  9  &   1    & 0    \\ 
$f_{16}\dotfill$ & 2510.0 & 398.41 &  6.8 &  8  &   1    & ?    \\ 
$f_{17}\dotfill$ & 2515.2 & 397.59 &  3.2 &  8  &   1    & ?    \\ 
 $f_{5}+f_{6}\dotfill$ & 3049.4 & 327.93 &  2.2 &$\cdots$&$\cdots$&$\cdots$\\
 $f_{5}+f_{8}\dotfill$ & 3130.9 & 319.40 &  3.7 &$\cdots$&$\cdots$&$\cdots$\\
 $f_{5}+f_{9}\dotfill$ & 3141.4 & 318.33 &  3.5 &$\cdots$&$\cdots$&$\cdots$\\
 $f_{8}+f_{8}\dotfill$ & 3224.2 & 310.15 &  2.4 &$\cdots$&$\cdots$&$\cdots$\\
$f_{5}+f_{10}\dotfill$ & 3248.6 & 307.83 &  2.2 &$\cdots$&$\cdots$&$\cdots$\\
$f_{8}+f_{10}\dotfill$ & 3341.9 & 299.23 &  3.2 &$\cdots$&$\cdots$&$\cdots$\\
$f_{9}+f_{12}\dotfill$ & 3585.9 & 278.87 &  2.9 &$\cdots$&$\cdots$&$\cdots$\\
$f_{5}+f_{15}\dotfill$ & 3836.5 & 260.65 &  2.4 &$\cdots$&$\cdots$&$\cdots$\\
$f_{9}+f_{15}\dotfill$ & 3940.4 & 253.78 &  3.3 &$\cdots$&$\cdots$&$\cdots$\\
$f_{5}+f_{16}\dotfill$ & 4028.8 & 248.22 &  2.0 &$\cdots$&$\cdots$&$\cdots$\\
$f_{8}+f_{16}\ldots$   & 4122.1 & 242.59 &  2.1 &$\cdots$&$\cdots$&$\cdots$\\
\hline
\end{tabular}
\end{center}
\end{table}


\begin{table}
\begin{center}
\caption{Optimal model parameters for CBS~114 and GD~358.\label{tab3}}
\begin{tabular}{lcrcrcl}
\hline
Parameter                      && CBS~114 && GD~358  && Error      \\ 
\hline
$T_{\rm eff}$~(K)$\dotfill$    && 25\,800 && 23\,100 && $\pm100$   \\
$M_*\ (M_{\odot})\dotfill$     && 0.630   && 0.630   && $\pm0.005$ \\
$\log(M_{\rm env}/M_*)\ldots$  && $-$2.42 && $-$2.92 && $\pm0.02$  \\
$\log(M_{\rm He}/M_*)\dotfill$ && $-$5.96 && $-$5.90 && $\pm0.02$  \\
$\sigma_{\rm P}$~(s)$\dotfill$ && 2.33    && 2.26    && $\cdots$   \\
\hline
\end{tabular}
\end{center}
\end{table}


\section{Model Fitting}

To investigate whether these new observations can fit comfortably with the
predictions of diffusion theory, we repeated the global model-fitting
procedure described by \cite{mmk03}. We used an updated version of the
code that incorporates the OPAL radiative opacities \citep{ir96} rather
than the older LAO data \citep{hue77}, which are known to produce
systematic errors in the derived temperatures \citep{fb94}. The fitting
procedure uses a parallel genetic algorithm \citep{mc03} to minimize the
root-mean-square residuals between the observed and calculated periods
($\sigma_{\rm P}$) for models with effective temperatures ($T_{\rm eff}$)
between 20\,000 and 30\,000~K, and stellar masses ($M_*$) between 0.45 and
0.95 $M_\odot$. It allows the base of the uniform He/C envelope to be
located at an outer mass fraction $\log(M_{\rm env}/M_*)$ between $-2.0$
and $-4.0$. The base of the pure He surface layer can assume values of
$\log(M_{\rm He}/M_*)$ between $-5.0$ and $-7.0$. To ensure a smooth 
transition between the self-consistent cores and the static envelopes in 
our models, we fixed the core composition to be pure C. We cannot 
presently include oxygen in the envelopes of our models.

We applied this fitting procedure to our new observations of CBS~114 and
to the WET observations of GD~358 from 1990 \citep{win94} for comparison.  
For CBS~114 we assumed that each single mode had an azimuthal order $m=0$
\citep[see][]{met03a}, and we used the central frequency for each of the
observed triplets. For the two doublets, we used the average of the two
observed frequencies. The resulting sets of optimal model parameters for
these two stars are shown in Table~\ref{tab3} with internal errors for each
parameter set by the resolution of our search. The theoretical pulsation
frequencies for CBS~114 are shown in Fig.~\ref{fig3} as solid squares
where they match an observed frequency, and the corresponding mode
identifications are listed in Table~\ref{tab2}. For reference, the open
squares in Fig.~\ref{fig3} show additional model frequencies with no
clearly detected counterpart.

The spectroscopic temperature determinations of \cite{bea99} suggest that
CBS~114 could be as much as 1500~K hotter than GD~358. Our model-fitting
results suggest a larger temperature difference, placing CBS~114 at
$2700\pm200$~K hotter than GD~358. On an absolute scale, the temperature
and mass of our fit to CBS~114 are consistent with the spectroscopic
values, assuming that any traces of H in the atmosphere are small. Our fit
to GD~358 is consistent with the spectroscopic mass, but our temperature
determination is slightly low. Our value is also low compared to a more
recent and precise determination using UV spectroscopy from the Hubble
Space Telescope \citep[$24\,100\pm400$~K;][]{cas05}.

Recent time-dependent diffusion calculations by \cite{ac04} suggest that
for a given stellar mass and envelope thickness, the pure He surface layer
should thicken by $\sim$0.15~dex between the spectroscopic temperatures of
CBS~114 and GD~358. We find a marginal thickening of only
$0.06\pm0.04$~dex. However, diffusion models with more massive envelopes
lead to thicker surface He layers more quickly as the star cools, since
there is a larger reservoir of He to draw from. Our fit to CBS~114 has a
slightly larger total envelope mass, which would tend to diminish the
expected difference between the thickness of the surface He layers for the
two stars to $\sim$0.12 dex. The limited grid of published diffusion
models currently makes more quantitative comparisons impossible.

\section{Discussion}

Our new dual-site observations of the pulsating DB white dwarf CBS~114
have finally revealed enough independent modes to bring the model-fitting
for this star to the same level of reliability as for GD~358. We used our
updated double-layered envelope models to match the new observations of
CBS~114, and the archival observations of GD~358---fitting 11 pulsation
modes in each star. The resulting pair of optimal models have pure He
surface layers and total envelope masses that qualitatively agree with our
expectations from diffusion theory. This provides new asteroseismic
evidence supporting one of the central assumptions of spectral evolution
theory, linking the DB white dwarfs to PG~1159 stars.

In addition to the new observational constraints for CBS~114, we improved
our analysis by incorporating the OPAL radiative opacities into the
models. Our results include higher temperatures and lower masses than the
previous fits by \cite{mmk03}, leading to better agreement with the
spectroscopic estimates for both stars. The optimal locations for the base
of the pure He surface layer and the uniform He/C envelope are
systematically deeper for GD~358 relative to the previous fit. For CBS~114
the fit is now consistent with the envelope mass expected from simulations
of carbon dredge-up in DQ stars \citep[see][]{pel86,fb02}, and with the He
layer mass expected from time-dependent diffusion calculations
\citep{ac04}.

These results have now met the challenge posed by \cite{met03}, to fit the
observations of more than one DBV star and demonstrate that they are both
consistent with the same underlying physical process. In that paper, the
relevant process was the nuclear burning history during the red giant
phase leading to a C/O ratio in the core that could help to constrain the
rate of the $^{12}{\rm C}(\alpha,\gamma)^{16}{\rm O}$ reaction. In the
present case, the relevant process is diffusion in the white dwarf
envelopes leading to a gradually thickening pure He surface layer as the
star cools. In effect, each of these studies has concentrated on an
isolated structure in the stellar interior, ignoring the contribution of
other possible features. Regardless of whether the asteroseismic signal
originates from the core or the envelope, it is now clear that the
relevant structure (C/O profile or He surface layer) is very nearly the
same in the two stars. The better {\it absolute} quality of the fits for
GD~358 and CBS~114 in \cite{met03} suggest that the C/O profile is the
more important structure from an asteroseismic standpoint. Indeed, this is
consistent with the finding by \cite{mmw03} that the pure He surface layer
has a smaller effect on the pulsation modes than the uniform He/C
envelope, which is itself secondary in importance to the C/O profile in
the core.

Of course, there are sound physical reasons to expect complicated
structures in both the core and the envelope of real white dwarfs. We have
demonstrated that analyses of two completely independent sets of pulsation
data with two physically distinct classes of models can both lead to
reasonably self-consistent interpretations of the corresponding internal
structures. Our challenge is now to synthesize hybrid models that can fit
the core and envelope structure simultaneously. It is possible that such
models will exhaust the information content of the 11 pulsation modes we
have observed in these two stars. If so, our only recourse is to detect
additional modes in GD~358 and CBS~114 and to discover additional DBV
stars with rich pulsation spectra. Both options will probably require
multi-site campaigns using 4\,m class telescopes since the existing
observations are already the result of campaigns using 1-2\,m telescopes,
and the newly discovered DBV stars from the Sloan Digital Sky Survey are
significantly fainter \citep{nit05}. Even so, the future continues to look
bright for white dwarf asteroseismology.

\begin{acknowledgements}
We would like to thank Steve Kawaler and Mike Montgomery for their 
comments and suggestions. This work was partially supported by the 
Smithsonian Institution through a CfA Postdoctoral Fellowship, and by 
the National Science Foundation through an Astronomy \& Astrophysics 
Postdoctoral Fellowship under award AST-0401441. Computational resources 
were provided by White Dwarf Research Corporation through grants from the 
Fund for Astrophysical Research and from the American Astronomical Society.
\end{acknowledgements}

\bibliographystyle{aa}       
\bibliography{cbs}

\end{document}